A new insight into the secondary path modeling problem in active noise control

Meiling Hu

Key Laboratory of Modern Acoustics, Institute of Acoustics, Nanjing University, Nanjing 210093, China.

Jun Wang

Key Laboratory of Modern Acoustics, Institute of Acoustics, Nanjing University, Nanjing 210093, China.

Jinpei Xue

Key Laboratory of Modern Acoustics, Institute of Acoustics, Nanjing University, Nanjing 210093, China.

Jing Lu[a)]

Key Laboratory of Modern Acoustics, Institute of Acoustics, Nanjing University, Nanjing 210093, China.

[a)] Electronic mail: lujing@nju.edu.cn

Running title: New insight into secondary path modeling

**Abstract**: The close relationship between the feedforward ANC system and the stereo acoustic echo cancellation system is revealed in this paper. Accordingly, the convergence behavior of the ANC system can be analyzed by investigating the joint auto-correlation matrix of the reference and the filtered reference signal. It is proved that the straightforward secondary path modeling can be carried out without the injection of any additive noise as long as the control filter is of a sufficiently long length. Furthermore, by taking advantage of the time-varying characteristic of the control filter, effective modeling of the secondary path can be even achieved without any restriction on the control filter length.

# I. INTRODUCTION

Active noise control (ANC) system introduces secondary sources to interfere destructively with unwanted noise and acts as an efficient alternative to the bulky passive noise control techniques for solving low-frequency noise problems(Bai et al., 2018; Behera et al., 2017; Cheer et al., 2018; Elliott, 2000; Hansen et al., 2012; Lu et al., 2003). The modeling of the secondary path, the transfer function between the control source and the error sensor, plays a vital role in the ANC system. Usually, an effective and efficient estimate of the secondary paths largely determines the performance of the ANC system (Elliott, 2000; Hansen et al., 2012; Kajikawa et al., 2012).

The normal way to estimate the secondary path is to launch a separate modeling process, usually called the off-line modeling, before the control process (Kajikawa et al., 2012). The online modeling method, with the injection of additive noise, has been proposed and optimized (Akhtar et al., 2006; Ming et al., 2001) to efficiently track the variation of the secondary path. To avoid the disadvantages caused by the additive noise, the simultaneous equation method (SEM) (Fujii et al., 2002; Jin et al., 2007) and overall modeling algorithm (OMA) were proposed (Kuo and Wang, 1992). The SEM aims to estimate the secondary path by algebraically solving the equations mixing the secondary path and an overall path related to both the secondary path and the primary path. The OMA is much simpler, but it needs to stop the control process and estimate the secondary path separately by setting the control filter as a pure delay. The unresolved meaningful question remains: under what condition is it possible to model the secondary path using the output of the control filter directly?

In this paper, the close relationship between the feedforward ANC system and the stereo acoustic echo cancellation (SAEC) system (Benesty et al., 1998; Gänsler and Benesty, 2000; Makino, 2001) is revealed, and it is found that the secondary path can be regarded as an echo transfer path and is possible to be identified directly using the output of the control filter. By investigating the joint auto-correlation matrix of the reference signal and the filtered reference signal, it is proved that only the restriction on the length of the control filter is required to guarantee the convergence of the secondary path modeling process. Furthermore, even this restriction can be neglected by taking advantage of the variation of the control filter.

## II. THEORY

### A. Relationship between ANC and SAEC

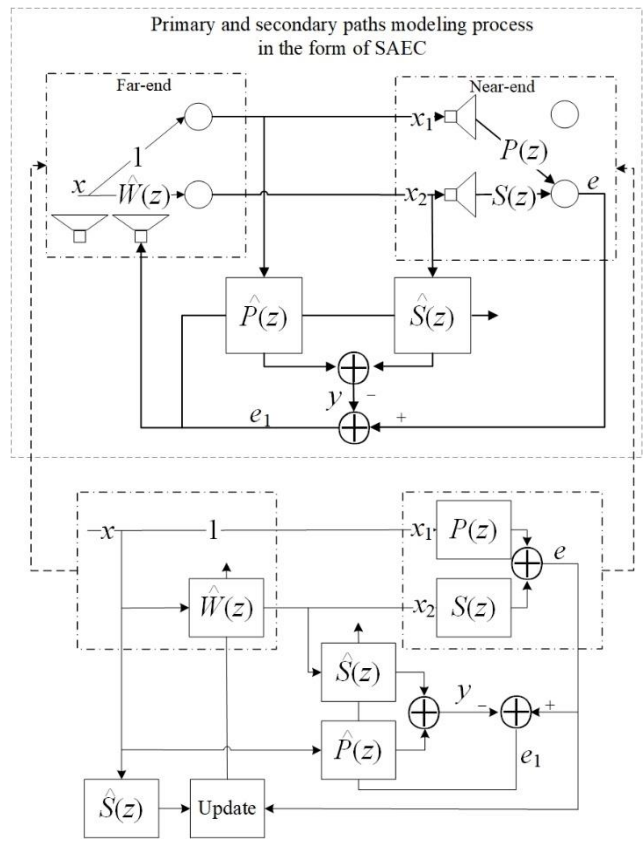

Fig. 1. The equivalence between the ANC modeling and control process and SAEC.

When the output of the control filter is utilized directly to model the secondary path while launching the control process simultaneously, the feedforward ANC system is closely related to a stereo acoustic echo cancellation system as shown in Fig. 1, where $P(z)$ represents the primary path between the reference sensor and the error sensor, $S(z)$ the secondary path between the control source and the error sensor and $W(z)$ the control filter. Described in the SAEC configuration, $P(z)$ and $S(z)$ can be regarded as the transfer functions between the stereo outputs and one of the microphones in the near-end, while $W(z)$ can be regarded as the transfer function between the speaker and one of the microphones in the far-end. The transfer function between the speaker and the other microphone in the far-end is just a constant 1. For ANC system, $\mathbf{x}_1$ is the same as the reference signal $\mathbf{x}$, while for SAEC system, $\mathbf{x}$ is the source signal from the speaker in the far-end and $\mathbf{x}_1$ is the output from one of the loudspeakers. Similarly, $\mathbf{x}_2$ acts as either the output of the control filter in ANC system or the output of the other loudspeaker in SAEC system, $e$ acts as either the error signal in ANC system or the signal captured by the microphone in the near-end in SAEC system, and $y$ acts as either the summation of the output of both the primary modeling filter and the secondary path modeling filter in ANC system or the summation of the output of the two adaptive filters in SAEC system. When the secondary path modeling and the control process are launched at the same time, the non-uniqueness problem (Benesty et al., 1998; Makino, 2001) exists in ANC system due to its close relationship with the SAEC system. Thus, it is meaningful to investigate under what condition the modeling process can be effectively conducted.

**B. Convergence analysis with fixed control filter**

In the following analysis, $\mathbf{w}$, $\mathbf{p}$ and $\mathbf{s}$ denote $\hat{W}(z), \hat{P}(z), \hat{S}(z)$ in time domain respectively. Define $\mathbf{w} = [w(1)\ w(2)\ \cdots\ w(N)]^T$, $\mathbf{p} = [p(1)\ p(2)\ \cdots\ p(L)]^T$ and $\mathbf{s} = [s(1)\ s(2)\ \cdots\ s(M)]^T$ as the control filter, the primary modeling filter and the secondary modeling filter with length $N$, $L$ and $M$ respectively and the superscript T denotes the transpose operation.

As shown in Fig. 1, $x_1$ is the same as input $x$ and $x_2$ is the output of the control filter. Define

$$\mathbf{x}_1(n) = [x_1(n) \ x_1(n-1) \ \cdots \ x_1(n-T+1)]^T$$
$$\mathbf{x}_2(n) = [x_2(n) \ x_2(n-1) \ \cdots \ x_2(n-T+1)]^T, \quad (1)$$

and

$$\mathbf{X}_1(n) = \begin{bmatrix} \mathbf{x}_1^T(n) \\ \mathbf{x}_1^T(n-1) \\ \vdots \\ \mathbf{x}_1^T(n-L+1) \end{bmatrix}, \mathbf{X}_2(n) = \begin{bmatrix} \mathbf{x}_2^T(n) \\ \mathbf{x}_2^T(n-1) \\ \vdots \\ \mathbf{x}_2^T(n-M+1) \end{bmatrix}, \quad (2)$$

where $T \gg \max(L, M)$ and $n$ denotes the time index which will be omitted hereafter for clarity with a few annotated exceptions. $\mathbf{X}_1$ and $\mathbf{X}_2$ are with the dimension of $L \times T$ and $M \times T$ respectively. In this paper, the noise is assumed to be a common stationary process and not a line spectral process, so that both these matrices are full row rank (Vaidyanathan, 2007). The output of the modeling filters, as shown in Fig. 1, can be expressed as

$$\mathbf{y} = \mathbf{X}_1^T \mathbf{p} + \mathbf{X}_2^T \mathbf{s}, \quad (3)$$

where $\mathbf{y}$ is column vector with length of $T$. Correspondingly, the Wiener equation is (Farhang-Boroujeny, 2013)

$$\mathbf{R}_x \begin{bmatrix} \mathbf{p} \\ \mathbf{s} \end{bmatrix} = \mathbf{r}_{xe}, \quad (4)$$

where $\mathbf{r}_{xe}$ denotes the joint cross-correlation vector calculated as

$$\mathbf{r}_{xe} = \frac{1}{T} \begin{bmatrix} \mathbf{X}_1 \\ \mathbf{X}_2 \end{bmatrix} \mathbf{e}, \quad (5)$$

with $\mathbf{e} = [e(1) \ e(2) \ \cdots \ e(T)]^T$ the desired output of the sum of the primary and secondary path. $\mathbf{R}_x$ is the joint auto-correlation matrix that can be calculated as

$$\mathbf{R}_x = \begin{bmatrix} \mathbf{R}_{x_1} & \mathbf{R}_{x_1,x_2} \\ \mathbf{R}_{x_1,x_2}^T & \mathbf{R}_{x_2} \end{bmatrix}, \quad (6)$$

with

$$\mathbf{R}_{x_1} = \frac{1}{T}\mathbf{X}_1\mathbf{X}_1^{\mathrm{T}}$$

$$\mathbf{R}_{x_2} = \frac{1}{T}\mathbf{X}_2\mathbf{X}_2^{\mathrm{T}} \ . \tag{7}$$

$$\mathbf{R}_{x_1,x_2} = \frac{1}{T}\mathbf{X}_1\mathbf{X}_2^{\mathrm{T}}$$

$\mathbf{R}_{x_1}$ and $\mathbf{R}_{x_2}$ are full rank unless the noise is a line spectral process (Vaidyanathan, 2007).

Similar to the analysis of the SAEC system (Makino, 2001), $x_1$ and $x_2$ are highly correlated signals from the same source, so that $\mathbf{R}_x$ is highly likely to be rank-deficient, resulting in non-unique solution of **p** and **s**. To investigate the rank of $\mathbf{R}_x$, Eq. (6) can be transformed into a block diagonal matrix with the same rank as

$$\begin{bmatrix} \mathbf{I} & \mathbf{0} \\ -\mathbf{R}_{x_1,x_2}^{\mathrm{T}}\mathbf{R}_{x_1}^{-1} & \mathbf{I} \end{bmatrix} \mathbf{R}_x \begin{bmatrix} \mathbf{R}_{x_1}^{-1} & -\mathbf{R}_{x_1}^{-1}\mathbf{R}_{x_1,x_2}\mathbf{R}_{x_2}^{-1} \\ \mathbf{0} & \mathbf{R}_{x_2}^{-1} \end{bmatrix} = \begin{bmatrix} \mathbf{I} & \mathbf{0} \\ \mathbf{0} & -\mathbf{A}+\mathbf{I} \end{bmatrix}, \tag{8}$$

where

$$\mathbf{A} = \mathbf{R}_{x_1,x_2}^{\mathrm{T}}\mathbf{R}_{x_1}^{-1}\mathbf{R}_{x_1,x_2}\mathbf{R}_{x_2}^{-1} \ . \tag{9}$$

The rank of $\mathbf{R}_x$ are totally determined by the lower right block of the right side of Eq. (8), $-\mathbf{A}+\mathbf{I}$, where **A** is a matrix with the dimension of $M \times M$. Substitution of Eq. (7) into Eq. (9) yields

$$\mathbf{A} = \mathbf{X}_2\mathbf{P}_1\mathbf{X}_2^{\mathrm{T}}\left(\mathbf{X}_2\mathbf{X}_2^{\mathrm{T}}\right)^{-1}, \tag{10}$$

where

$$\mathbf{P}_1 = \mathbf{X}_1^{\mathrm{T}}\left(\mathbf{X}_1\mathbf{X}_1^{\mathrm{T}}\right)^{-1}\mathbf{X}_1 \ . \tag{11}$$

$\mathbf{P}_1$ represents the projection matrix that maps vectors onto the row space of $\mathbf{X}_1$ (Strang, 2009). $\mathbf{P}_1$ is a symmetric matrix, with eigenvalues of either 0 or 1, and can be diagonalized as

$$\mathbf{P}_1 = \mathbf{Q}\mathbf{\Lambda}\mathbf{Q}^{-1} = \mathbf{Q}\mathbf{\Lambda}\mathbf{Q}^{\mathrm{T}}, \tag{12}$$

where **Q** is an orthonormal matrix and **Λ** is a diagonal matrix with elements of either 0 or 1. The identity matrix can be decomposed as

$$\mathbf{I} = \mathbf{X}_2\mathbf{Q}\mathbf{Q}^{\mathrm{T}}\mathbf{X}_2^{\mathrm{T}}\left(\mathbf{X}_2\mathbf{X}_2^{\mathrm{T}}\right)^{-1}. \tag{13}$$

Combining Eq. (13) with Eq. (10) yields

$$-\mathbf{A}+\mathbf{I} = \mathbf{X}_2\mathbf{P}_2\mathbf{X}_2^\mathrm{T}\left(\mathbf{X}_2\mathbf{X}_2^\mathrm{T}\right)^{-1}, \qquad (14)$$

where

$$\mathbf{P}_2 = \mathbf{Q}(-\mathbf{\Lambda}+\mathbf{I})\mathbf{Q}^\mathrm{T} = \mathbf{L}\mathbf{L}^\mathrm{T} \qquad (15)$$

is the projection operator projecting onto the null space of $\mathbf{X}_1$ (Strang, 2009), and

$$\mathbf{L} = \mathbf{Q}(-\mathbf{\Lambda}+\mathbf{I}) \qquad (16)$$

can be regarded as the operator getting the coordinates of the projection of a vector in the null space of $\mathbf{X}_1$. It can be seen from Eq. (14) that the rank of $-\mathbf{A}+\mathbf{I}$ is determined by the rank of $\mathbf{X}_2\mathbf{P}_2\mathbf{X}_2^\mathrm{T}$ due to the rank-full property of $\mathbf{X}_2$. $\mathbf{X}_2\mathbf{P}_2\mathbf{X}_2^\mathrm{T}$ is a square matrix and can be expressed as $\mathbf{X}_2\mathbf{L}(\mathbf{X}_2\mathbf{L})^\mathrm{T}$ so that its rank is determined by $\mathbf{X}_2\mathbf{L}$. To analyze if $\mathbf{X}_2\mathbf{L}$ is full row rank, left multiplying it by a row vector $\mathbf{a}$ as $\mathbf{a}\mathbf{X}_2\mathbf{L}$. If there exists a non-zero $\mathbf{a}$ causing $\mathbf{a}\mathbf{X}_2$ to be in the row space of $\mathbf{X}_1$, then the projection operator $\mathbf{L}$ will get all-zero coordinates, which means $\mathbf{X}_2\mathbf{L}$ is not full row rank, resulting in a rank-deficient $-\mathbf{A}+\mathbf{I}$. On the other hand, if $\mathbf{a}\mathbf{X}_2$ is not in the row space of $\mathbf{X}_1$ for any non-zero vector $\mathbf{a}$, $\mathbf{X}_2\mathbf{L}$ will be guaranteed to be full row rank, leading to a rank-full $-\mathbf{A}+\mathbf{I}$.

From the above analysis, the rank of $\mathbf{R}_x$ is totally determined by whether or not there exists a non-zero $\mathbf{a}$ causing $\mathbf{a}\mathbf{X}_2$ to be in the row space of $\mathbf{X}_1$. From Fig. 1 it can be seen that $x_2$ is a filtered version of $x_1$, so that

$$\mathbf{a}\mathbf{X}_2 = \mathbf{a}\mathbf{W}\mathbf{X}_{1,*}, \qquad (17)$$

where

$$\mathbf{W} = \begin{bmatrix} w(1) & w(2) & \cdots & w(N) & 0 & \cdots & 0 \\ 0 & w(1) & w(2) & \cdots & w(N) & \ddots & \vdots \\ \vdots & 0 & \ddots & \cdots & \cdots & \ddots & 0 \\ 0 & \cdots & 0 & w(1) & w(2) & \cdots & w(N) \end{bmatrix} \qquad (18)$$

is a $M \times (N+M-1)$ matrix formed by $M$ rows of shifted control filter parameters and

$$\mathbf{X}_{1,*} = \begin{bmatrix} \mathbf{x}_1^T(n) \\ \mathbf{x}_1^T(n-1) \\ \vdots \\ \mathbf{x}_1^T(n-N-M+2) \end{bmatrix} \qquad (19)$$

is different from $\mathbf{X}_1$ only by the row number.

**Case 1**: $N+M-1 \leq L$. In this case, $\mathbf{X}_{1,*}$ is a part of $\mathbf{X}_1$, then $\mathbf{a}\mathbf{X}_2$ is always in the row space of $\mathbf{X}_1$.

**Case 2**: $N+M-1 > L$ and $N \leq L$. When $N+M-1 > L$, $\mathbf{a}\mathbf{X}_2$ can be further expressed as

$$\mathbf{a}\mathbf{X}_2 = \mathbf{a}\mathbf{W}\mathbf{X}_{1,*} = \mathbf{a}\mathbf{W}^{(1)}\mathbf{X}_{1,*}^{(1)} + \mathbf{a}\mathbf{W}^{(2)}\mathbf{X}_{1,*}^{(2)},$$

where

$$\mathbf{W} = \begin{bmatrix} \mathbf{W}^{(1)} & \mathbf{W}^{(2)} \end{bmatrix} \qquad (20)$$

with $\mathbf{W}^{(1)}$ the first $L$ columns of $\mathbf{W}$ and $\mathbf{W}^{(2)}$ the rest columns, and correspondingly $\mathbf{X}_{1,*}^{(1)}$ is exactly the same as $\mathbf{X}_1$ consisting of the first $L$ rows of $\mathbf{X}_{1,*}$ and $\mathbf{X}_{1,*}^{(2)}$ consists of the rest rows of $\mathbf{X}_{1,*}$. It can be seen that $\mathbf{a}\mathbf{X}_2$ is in the row space of $\mathbf{X}_1$ if $\mathbf{a}\mathbf{W}^{(2)}\mathbf{X}_{1,*}^{(2)} = \mathbf{0}$. The separation of $\mathbf{W}$ is indicated by the dotted line shown in Eq. (18). When $N \leq L$, the separation line between $\mathbf{W}^{(1)}$ and $\mathbf{W}^{(2)}$ either coincides with or on the right side of the dotted line, so that $\mathbf{W}^{(2)}$ has at least one all-zero row (the first row), resulting in deficient rank. Therefore, there always exists a non-zero $\mathbf{a}$ causing $\mathbf{a}\mathbf{W}^{(2)}\mathbf{X}_{1,*}^{(2)} = 0$, so that $\mathbf{a}\mathbf{X}_2$ is in the row space of $\mathbf{X}_1$.

**Case 3**: $N > L$. In this case the separation line indicated in Eq. (18) is on the left side of the dotted line, $\mathbf{W}^{(2)}$ is guaranteed to be full row rank if $w(L+1) \sim w(N)$ are not all zero. Then $\mathbf{a}\mathbf{X}_2$ is not in the row space of $\mathbf{X}_1$.

In summary, when the control filter length $N$ is set larger than that of the primary path modeling filter $L$, and the last $N–L$ coefficients of the control filter are not all zero, the joint auto-correlation matrix $\mathbf{R}_x$ is guaranteed to be rank-full, so that the non-unique problem can be avoided in the ANC modeling process. Note that the OMA scheme proposed in (Kuo and Wang, 1992) is a special case of this general conclusion, where the first $L$ coefficients of the control filter $w(1)\sim w(L)$ are deliberately set to zero to fulfill the modeling task.

**C. Convergence analysis with the time-varying control filter**

It has been noted that a time-varying filter between the two reference signals can help guarantee the convergence of the filters in the SAEC system (Ali, 1998; Sugiyama et al., 2001). When the secondary-path modeling and the control process are conducted simultaneously in the feedforward ANC system, the control filter acts exactly as a time-varying filter between $x_1$ and $x_2$. To investigate the behavior of the modeling process in this situation, define $\mathbf{p}_o = [p_o(1)\ p_o(2)\ \cdots\ p_o(L)]^T$ and $\mathbf{s}_o = [s_o(1)\ s_o(2)\ \cdots\ s_o(M)]^T$ as the optimal Wiener solution of the primary and the secondary path with length $L$ and $M$ respectively. Define $\mathbf{x}(n) = [x(n)\ x(n-1)\ \cdots\ x(n-K+1)]^T$ as the reference vector with $n$ the time index and $K$ the vector length where $K >= \max(N+M-1, L)$, then $e_1$ shown in Fig. 1 can be expressed as

$$e_1 = (\mathbf{p}_o - \mathbf{p})^T \mathbf{G}_1 \mathbf{x}(n) + (\mathbf{s}_o - \mathbf{s})^T \mathbf{W}\mathbf{G}_2 \mathbf{x}(n), \tag{21}$$

where $\mathbf{W}$ is in the form shown in Eq. (18) and

$$\mathbf{G}_1 = [\mathbf{I}\ \ \mathbf{0}]_{L \times K},\ \mathbf{G}_2 = [\mathbf{I}\ \ \mathbf{0}]_{(N+M-1) \times K} \tag{22}$$

with the subscript indicating the dimension of the matrix.

The minimum square error (MSE) is

$$E\{e_1^2\} = \mathbf{v}^T E\{\mathbf{xx}^T\} \mathbf{v}, \tag{23}$$

where

$$\mathbf{v} = \mathbf{G}_1^T (\mathbf{p}_o - \mathbf{p}) + \mathbf{G}_2^T \mathbf{W}^T (\mathbf{s}_o - \mathbf{s}). \tag{24}$$

Taking the derivative of the MSE with respect to $\mathbf{v}$ leads to

$$\frac{\partial E\{e_1^2\}}{\partial \mathbf{v}} = 2E\{\mathbf{xx}^T\} \mathbf{v}. \tag{25}$$

Note that $E\{\mathbf{xx}^T\}$ is full rank unless the noise is a line spectral process (Vaidyanathan, 2007), so $\frac{\partial E\{e_1^2\}}{\partial \mathbf{v}} = \mathbf{0}$ leads to $\mathbf{v} = \mathbf{0}$, i.e.

$$\mathbf{G}_1^T (\mathbf{p}_o - \mathbf{p}) + \mathbf{G}_2^T \mathbf{W}^T (\mathbf{s}_o - \mathbf{s}) = \mathbf{0}. \tag{26}$$

For a different control filter with corresponding matrix shown in Eq. (18) expressed as $\mathbf{W}_*$, it can be obtained that

$$\mathbf{G}_1^T (\mathbf{p}_o - \mathbf{p}) + \mathbf{G}_2^T \mathbf{W}_*^T (\mathbf{s}_o - \mathbf{s}) = \mathbf{0}. \tag{27}$$

Subtraction of Eq. (27) from Eq. (26) yields

$$\mathbf{G}_2^T \left( \mathbf{W}^T - \mathbf{W}_*^T \right)(\mathbf{s}_o - \mathbf{s}) = \mathbf{0}. \tag{28}$$

Note that $\mathbf{G}_2$ and $\mathbf{W} - \mathbf{W}_*$ are both full row rank, and accordingly their transposes are full column rank. Thus Eq. (28) indicates that $\mathbf{s} = \mathbf{s}_o$, which means that the modeling filter is guaranteed to converge to the optimal Wiener solution under time-varying control filter condition.

Note that the analysis above does not restrict the length of the control filter and the primary path, so that the restriction discussed in Sec. II.B can be further neglected in time-varying control filter condition.

**III. SIMULATIONS**

The primary and the secondary paths used in the simulations were measured from a duct with $L$

= 48 and $M = 48$, and the sampling rate is 1000 Hz. The control source was 30 cm away from the noise source while the error microphone was placed 15 cm downstream from the control source. The noise with pass-band between 200 Hz and 400 Hz is utilized in the simulations.

**A. Simulation with fixed control filter**

The first and the last coefficients of the control filter were initialized as 1 while all the others were set as zero. The control filter was fixed during the modeling process. The normalized least mean square (NLMS) algorithm was used because of its simplicity and wide application in ANC systems. The modeling results with different control filter length are compared in Fig. 2. It can be seen that when $N$ is set as 49, larger than $L$, the secondary path modeling result is exactly the same as the ideal one obtained by the off-line modeling method. However, when $N$ is set as 48, the same as $L$, the modeling result deviates from the ideal one, with modeling error of the first tap clearly seen. Furthermore, when $N$ is set as 24, significantly smaller than $L$, the modeling result differs considerably from the ideal one. The performance matches well with the analysis in Sec. II.B.

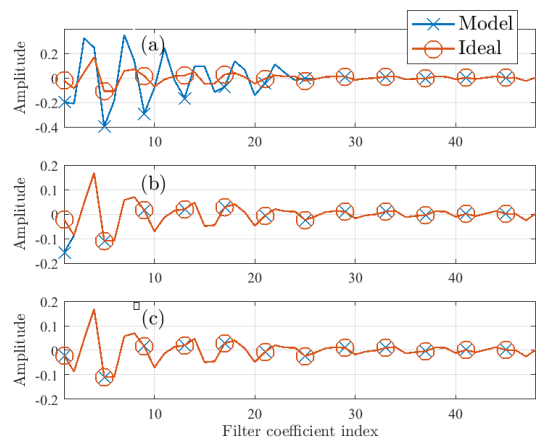

Fig. 2. (Color online) Modeling results of the secondary path with fixed control filter of different length. (a) $N = 24$, (b) $N = 48$, (c) $N = 49$.

**B. Simulation with time-varying control filter**

In this simulation, the length of control filter $N$ was set as 2 and the coefficients were alternated between [0 1] and [1 0] at each iteration. The secondary path modeling result with the NLMS algorithm is presented in Fig. 3. As expected from Sec. II.C, due to the time-varying characteristic of the control filter, the modeling result matches well with the ideal one even when the control filter has a much shorter length than that of the primary path.

During the ANC process, the control filter acts exactly as a time-varying filter, therefore it is natural to further investigate whether it is possible to launch the control and modeling process simultaneously. The control filter was initialized as the standard impulse response with length $N$ equal 48, and the NLMS algorithm is used for both the control filter updating and the modeling process as shown in Fig. 1. The secondary path modeling result and the noise attenuation performance are presented in Fig. 4. It can be seen that the noise is successfully attenuated while the modeling result deviates slightly from the ideal one, which can be attributed to the slow convergence of the NLMS algorithm. When the fast converging recursive least square (RLS) algorithm is utilized, the secondary path can be modeled precisely as shown in Fig. 5.

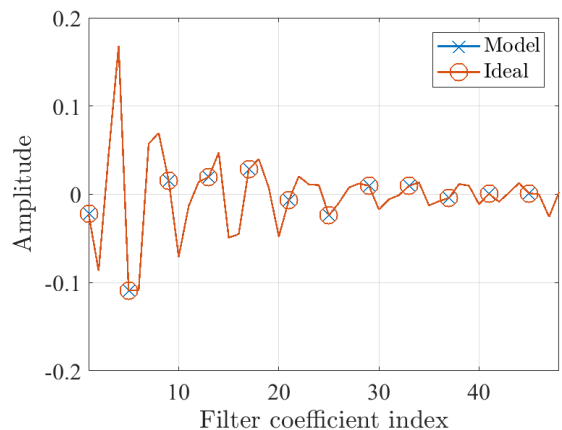

Fig. 3. (Color online) Modeling result of the secondary path with time-varying control filter.

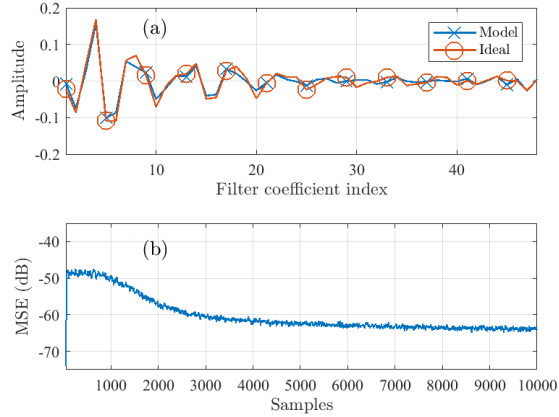

Fig. 4. (Color online) (a) Modeling result of the secondary path by NLMS algorithm; (b) Corresponding MSE variation.

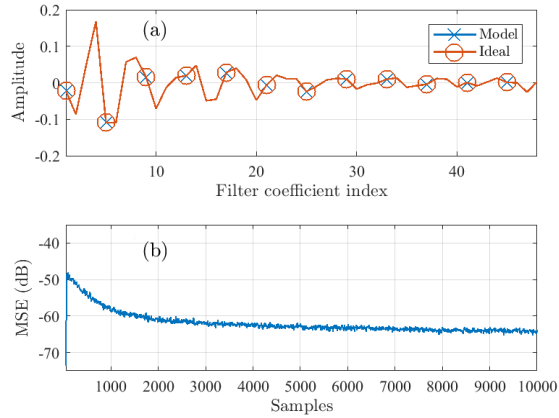

Fig. 5. (Color online) (a) Modeling result of the secondary path by RLS algorithm; (b) Corresponding MSE variation.

## IV. CONCLUSIONS

In this paper, the close relationship between the feedforward active control system and the stereo acoustic echo cancellation system is revealed, and the secondary path modeling process is analyzed by investigating the rank of the joint auto-correlation matrix. It is theoretically proved that the secondary path is guaranteed to converge to the optimal Wiener solution by just utilizing the control filter output as the reference signal if the control filter length is larger than that of the primary path. Furthermore, by taking advantage of the time-varying characteristic of the control filter, precise modeling of the secondary path can be achieved even without the restriction of filter length.

ACKNOWLEDGEMENTS

This work was supported by the National Natural Science Foundation of China (Grant No. 11874219).